

\catcode`\X=12\catcode`\@=11

\def\n@wcount{\alloc@0\count\countdef\insc@unt}
\def\n@wwrite{\alloc@7\write\chardef\sixt@@n}
\def\n@wread{\alloc@6\read\chardef\sixt@@n}
\def\r@s@t{\relax}\def\v@idline{\par}\def\@mputate#1/{#1}
\def\l@c@l#1X{\firstpart.#1}\def\gl@b@l#1X{#1}\def\t@d@l#1X{{}}

\def\crossrefs#1{\ifx\all#1\let\tr@ce=\all\else\def\tr@ce{#1,}\fi
   \n@wwrite\cit@tionsout\openout\cit@tionsout=\jobname.cit 
   \write\cit@tionsout{\tr@ce}\expandafter\setfl@gs\tr@ce,}
\def\setfl@gs#1,{\def\@{#1}\ifx\@\empty\let\next=\relax
   \else\let\next=\setfl@gs\expandafter\xdef
   \csname#1tr@cetrue\endcsname{}\fi\next}
\def\m@ketag#1#2{\expandafter\n@wcount\csname#2tagno\endcsname
     \csname#2tagno\endcsname=0\let\tail=\all\xdef\all{\tail#2,}
   \ifx#1\l@c@l\let\tail=\r@s@t\xdef\r@s@t{\csname#2tagno\endcsname=0\tail}\fi
   \expandafter\gdef\csname#2cite\endcsname##1{\expandafter
     \ifx\csname#2tag##1\endcsname\relax?\else\csname#2tag##1\endcsname\fi
     \expandafter\ifx\csname#2tr@cetrue\endcsname\relax\else
     \write\cit@tionsout{#2tag ##1 cited on page \folio.}\fi}
   \expandafter\gdef\csname#2page\endcsname##1{\expandafter
     \ifx\csname#2page##1\endcsname\relax?\else\csname#2page##1\endcsname\fi
     \expandafter\ifx\csname#2tr@cetrue\endcsname\relax\else
     \write\cit@tionsout{#2tag ##1 cited on page \folio.}\fi}
   \expandafter\gdef\csname#2tag\endcsname##1{\expandafter
      \ifx\csname#2check##1\endcsname\relax
      \expandafter\xdef\csname#2check##1\endcsname{}%
      \else\immediate\write16{Warning: #2tag ##1 used more than once.}\fi
      \multit@g{#1}{#2}##1/X%
      \write\t@gsout{#2tag ##1 assigned number \csname#2tag##1\endcsname\space
      on page \number\count0.}%
   \csname#2tag##1\endcsname}}
\def\multit@g#1#2#3/#4X{\def\t@mp{#4}\ifx\t@mp\empty%
      \global\advance\csname#2tagno\endcsname by 1 
      \expandafter\xdef\csname#2tag#3\endcsname
      {#1\number\csname#2tagno\endcsnameX}%
   \else\expandafter\ifx\csname#2last#3\endcsname\relax
      \expandafter\n@wcount\csname#2last#3\endcsname
      \global\advance\csname#2tagno\endcsname by 1 
      \expandafter\xdef\csname#2tag#3\endcsname
      {#1\number\csname#2tagno\endcsnameX}%
      \write\t@gsout{#2tag #3 assigned number \csname#2tag#3\endcsname\space
      on page \number\count0.}\fi
   \global\advance\csname#2last#3\endcsname by 1
   \def\t@mp{\expandafter\xdef\csname#2tag#3/}%
   \expandafter\t@mp\@mputate#4\endcsname
   {\csname#2tag#3\endcsname\lastpart{\csname#2last#3\endcsname}}\fi}
\def\t@gs#1{\def\all{}\m@ketag#1e\m@ketag#1s\m@ketag\gl@b@l c 
   \m@ketag\l@c@l b \m@ketag\t@d@l p
   \m@ketag\gl@b@l r \n@wread\t@gsin
   \openin\t@gsin=\jobname.tgs \re@der \closein\t@gsin
   \n@wwrite\t@gsout\openout\t@gsout=\jobname.tgs }
\outer\def\localtags{\t@gs\l@c@l}
\outer\def\globaltags{\t@gs\gl@b@l}
\outer\def\newlocaltag#1{\m@ketag\l@c@l{#1}}
\outer\def\newglobaltag#1{\m@ketag\gl@b@l{#1}}

\newif\ifpr@ 
\def\m@kecs #1tag #2 assigned number #3 on page #4.%
   {\expandafter\gdef\csname#1tag#2\endcsname{#3}
   \expandafter\gdef\csname#1page#2\endcsname{#4}
   \ifpr@\expandafter\xdef\csname#1check#2\endcsname{}\fi}
\def\re@der{\ifeof\t@gsin\let\next=\relax\else
   \read\t@gsin to\t@gline\ifx\t@gline\v@idline\else
   \expandafter\m@kecs \t@gline\fi\let \next=\re@der\fi\next}
\def\pretags#1{\pr@true\pret@gs#1,,}
\def\pret@gs#1,{\def\@{#1}\ifx\@\empty\let\n@xtfile=\relax
   \else\let\n@xtfile=\pret@gs \openin\t@gsin=#1.tgs \message{#1} \re@der 
   \closein\t@gsin\fi \n@xtfile}

\newcount\sectno\sectno=0\newcount\subsectno\subsectno=0
\newif\ifultr@local \def\ultralocal{\ultr@localtrue}
\def\firstpart{\number\sectno}
\def\lastpart#1{\ifcase#1 \or a\or b\or c\or d\or e\or f\or g\or h\or 
   i\or k\or l\or m\or n\or o\or p\or q\or r\or s\or t\or u\or v\or w\or 
   x\or y\or z \fi}
\def\closeup{\vskip-\bigskipamount}
\def\resetall{\global\advance\sectno by 1\subsectno=0
   \gdef\firstpart{\number\sectno}\r@s@t}
\def\resetsub{\global\advance\subsectno by 1
   \gdef\firstpart{\number\sectno.\number\subsectno}\r@s@t}
\def\newsection#1#2\par{\resetall\vskip0pt plus.1\vsize\penalty-250
   \vskip0pt plus-.1\vsize\bigskip\bigskip
   \message{#2}\leftline{\bf \ctag{#1}.\ \vtop{\hsize6.2truein
  \raggedright\noindent\hangafter1\hangindent20pt #2}}\nobreak\bigskip}
\def\subsection#1#2\par{\ifultr@local\resetsub\fi
   \vskip0pt plus.1\vsize\penalty-250\vskip0pt plus-.1\vsize
   \bigskip\smallskip\message{#2}\leftline{\bf\btag{#1}#2}\nobreak\medskip}

\def\(#1){\edef\dot@g{\ifmmode\ifinner(\hbox{\rm \noexpand\etag{#1}})
   \else\noexpand\eqno(\hbox{\noexpand\etag{#1}})\fi
   \else(\noexpand\ecite{#1})\fi}\dot@g}

\newif\ifbr@ck
\def\eat#1{}
\def\[#1]{\br@cktrue[\br@cket#1'X]}
\def\br@cket#1'#2X{\def\temp{#2}\ifx\temp\empty\let\next\eat
   \else\let\next\br@cket\fi
   \ifbr@ck\br@ckfalse\br@ck@t#1,X\else\br@cktrue#1\fi\next#2X}
\def\br@ck@t#1,#2X{\def\temp{#2}\ifx\temp\empty\let\neext\eat
   \else\let\neext\br@ck@t\def\temp{,}\fi
   \def\teemp{#1}\ifx\teemp\empty\else\rcite{#1}\fi\temp\neext#2X}
\def\resetbr@cket{\gdef\[##1]{[\rtag{##1}]}}
\def\references\par{\resetall\resetbr@cket\vskip0pt plus.2\vsize\penalty-250
   \vskip0pt plus-.3\vsize\bigskip\bigskip
   \message{References}\leftline{\bf References}\nobreak\bigskip}

\catcode`\X=11 \catcode`\@=12


\magnification=\magstep1
\input epsf

\localtags
\newglobaltag{f}

\def\thm#1{\medskip\noindent{\bf Theorem \stag{#1}: }}
\def\lthm#1#2{\medskip\noindent{\bf Theorem \stag{#1} {\rm #2}: }}
\def\lem#1{\medskip\noindent{\bf Lemma \stag{#1}: }}
\def\rmk#1{\medskip\noindent{\bf Remark \stag{#1}: }}
\def\ex#1{\medskip\noindent{\bf Example \stag{#1}: }}
\def\fig#1{\noindent{\bf Figure \ftag{#1}: }}
\def\llem#1#2{\medskip\noindent{\bf Lemma \stag{#1} {\rm #2}: }}
\def\csec#1{Section~\ccite{#1}}
\def\cssec#1{Section~\bcite{#1}}
\def\cor#1{\medskip\noindent{\bf Corollary \stag{#1}: }}

\def\cthm#1{Theorem~\scite{#1}}
\def\clem#1{Lemma~\scite{#1}}
\def\crmk#1{Remark~\scite{#1}}
\def\cex#1{Example~\scite{#1}}
\def\cfig#1{Figure~\fcite{#1}}

\def\ccor#1{Corollary~\scite{#1}}
\def\proof{\medskip\noindent{\sl Proof: }}
\def\proofof#1{\medskip\noindent{\sl Proof of #1: }}
\def\qed{\nobreak\kern 1em \vrule height .5em width .5em depth 0em}

\def\Re{\mathop{\rm Re}}\def\Im{\mathop{\rm Im}}
\input amssym.tex
\def\bbz{{\Bbb Z}}\def\bbr{{\Bbb R}}
\def\bbc{\Bbb C}\def\bbl{\Bbb L}
\def\usigma{\underline\sigma}

\def\ds{\displaystyle}

\def \today {\ifcase \month \or January\or February\or  
  March\or April\or May\or June\or July\or August\or 
  September\or October\or November\or December\fi\   
  \number \day, \number \year}                       
\def\l{l}
\def\G{{\cal G}}

 at 14.4 truept
\nopagenumbers

\null\vfill\centerline{\bf LOCATION OF THE LEE-YANG ZEROS AND ABSENCE OF}
 \medskip
 \centerline{\bf PHASE TRANSITIONS IN SOME ISING SPIN SYSTEMS*}

 \bigskip\bigskip
 \centerline{Joel L. Lebowitz${}^{1,2}$, 
  David Ruelle${}^{1,3}$, and Eugene R. Speer${}^1$}

\bigskip
\centerline{${}^1$Department of Mathematics, Rutgers
University,}
\centerline{Piscataway, New Jersey 08854 USA}
 \smallskip
 \centerline{${}^2$Department of Physics, Rutgers
University,}
\centerline{Piscataway, New Jersey 08854 USA}
 \smallskip
 \centerline{${}^3$IHES, 91440 Bures sur Yvette, France}
 \bigskip
 \centerline{Submitted April 9, 2012}
 \smallskip
 \centerline{Revised version June 29, 2012}
 \bigskip\bigskip
\centerline{\bf Abstract}
 \medskip
We consider a class of Ising spin systems on a set $\Lambda$ of sites.  The
sites are grouped into units with the property that each site belongs to
either one or two units, and the total internal energy of the system is the
sum of the energies of the individual units, which in turn depend only on
the number of up spins in the unit.  We show that under suitable conditions
on these interactions none of the $|\Lambda|$ Lee-Yang zeros in the complex
$z=e^{2\beta h}$ plane, where $\beta$ is the inverse temperature and $h$
the uniform magnetic field, touch the positive real axis, at least for
large values of $\beta$.  In some cases one obtains, in an appropriately
taken $\beta\nearrow\infty$ limit, a gas of hard objects on a set
$\Lambda'$; the fugacity for the limiting system is a rescaling of $z$ and
the Lee-Yang zeros of the new partition function also avoid the positive
real axis.  For certain forms of the energies of the individual units the
Lee-Yang zeros of both the finite- and zero-temperature systems lie on the
negative real axis for all $\beta$.  One zero-temperature limit of this
type, for example, is a monomer-dimer system; our results thus generalize,
to finite $\beta$, a well-known result of Heilmann and Lieb that the
Lee-Yang zeros of monomer-dimer systems are real and negative.
 \vfill
 
\noindent
*Dedicated to Elliott Lieb on the occasion of his eightieth birthday, in
friendship and admiration. 
\vfill\eject
\pageno1
\footline{\hfill\rm \folio\hfill}

\newsection{intro} Introduction

 We consider a system of Ising spins on a finite set $\Lambda$ of sites; we
often think of $\Lambda$ as a subset of some lattice $\bbl$.  Writing
$\usigma=(\sigma_i)_{i\in\Lambda}$, with $\sigma_i=\pm1$, for a spin
configuration, we let $N=N(\usigma)=\sum_{i\in\Lambda} (1+\sigma_i)/2$ be
the total number of up spins.  We will sometimes think of this system as a
lattice gas in which $\eta_i=(1+\sigma_i)/2$ is the indicator of a particle
at site $i$; $N$ is then the total number of particles in the system and
$N/|\Lambda|$, with $|\Lambda|$ the number of sites, the average density
$\rho$. The average magnetization per site is
$m=|\Lambda|^{-1}\sum\sigma_i =|\Lambda|^{-1}(2N-|\Lambda|)=2\rho-1$.  The
thermodynamic properties of this system are determined \[Ruellebook] by the
partition function
 $$ Z_\Lambda(z,\beta)=\sum_{\underline\sigma:\Lambda\to\pm1}
	z^{N(\underline\sigma)}e^{-\beta U(\underline\sigma)}, \(zdef) $$
where $U(\usigma)$ is the internal energy of the configuration $\usigma$,
$\beta$ is the inverse temperature, and $z$ is the magnetic fugacity, that
is, $z=e^{2\beta h}$ with $h$ the magnetic field.  $Z_\Lambda$ is a
polynomial in $z$  of order $|\Lambda|$, with positive coefficients.

The zeros in the complex fugacity plane of $Z_\Lambda(z,\beta)$, usually
called Lee-Yang zeros, have been of interest since the original studies of
Yang and Lee \[YL] and Lee and Yang \[LY].  For finite systems none of the
$|\Lambda|$ zeros can lie on the physically relevant positive real axis.
But when $\Lambda$ is a subset of some periodic lattice $\bbl$ and
$U(\usigma)$ is the restriction to $\Lambda$ of a translation invariant
energy (with some boundary conditions), so that we may speak of the
thermodynamic limit $\Lambda\nearrow\bbl$, the zeros can in this limit
approach the real axis, signaling (typically) the existence of a phase
transition in the model \[YL].  The nature of the phase transition depends
on the manner in which the zeros approach the positive $z$-axis as $\beta$
or other parameters in $U$ are changed.  Speaking loosely, there will be a
discontinuity in the magnetization per site, that is, a first order
transition, at a value $H$ of the magnetic field if the density of zeros on
the real axis at $z=e^{2\beta H}$ is nonzero, and a higher order transition
if there is a nonzero density arbitrarily close to this point \[YL].

In their second paper \[LY], Lee and Yang proved that for the
Ising model with ferromagnetic pair interactions, that is, for
 $$U(\usigma)=-\sum_{\{i,j\}\in\Lambda,\;i\ne j}J_{ij}\sigma_i\sigma_j, 
   \(ising)$$
  with all $J_{ij}\ge0$, all the zeros of $Z_\Lambda$ lie on the unit
circle $|z|=1$.  Consequently, the only possible thermodynamic phase
transition in this system takes place at $z=1$ or $h=0$.  The Lee-Yang
theorem has been extended in many ways to a variety of classical and
quantum systems; see \[SF,BBCKK] for reviews.  One can also prove in
many cases that there is indeed a first order phase transition at
sufficiently large $\beta$, so that the zeros must have a nonzero density
at $z=1$ in the thermodynamic limit.

Much less is known rigorously for general spin systems in which the zeros
do not lie on the unit circle.  This has led to numerical studies of these
zeros for $\Lambda$ a subset of a lattice $\bbl$; in particular, the cases
in which $\bbl$ is either $\bbz^2$ or the planar triangular lattice, and in
which the internal energy is given by \(ising) with uniformly
antiferromagnetic nearest-neighbor interactions, that is, with
$J_{ij}=J\delta_{|i-j|,1}$, $J<0$, have been investigated extensively
\[Kim,HK,LR1].  These systems can be proven to undergo phase transitions in
the thermodynamic limit $\Lambda\nearrow\bbl$, for large values of $\beta$,
at nonzero values of $h$ \[D,DKS,Sin].  This implies that the zeros of their
partition functions must converge to the real axis at some point
$z(\beta)\ne1$.  More recently there have also been results for systems in
which the zeros lie on the unit circle for large $\beta$ but not for small
$\beta$ \[LR].  In some cases they touch the real axis, either for finite
$\beta$ or in the limit $\beta\nearrow\infty$.

There have also been many studies of  the Lee-Yang zeros of the
grand canonical partition function for general interacting particle systems
on lattices or in the continuum.  Of particular interest to us is the case
of ``hard'' interactions, in which for every particle configuration
$\eta$ either $U(\eta)=0$ or $U(\eta)=\infty$; put another way, some
configurations are forbidden, while all others have no internal energy.
Systems with such interactions can often be obtained as a suitable
$\beta\to\infty$ limit of \(ising).  In many interesting cases one may then
think of the model as a system of particles (which may or may not
correspond to the original particles) with fixed shapes, like dimers,
diamonds, or hexagons, which cover more than one lattice site and which
cannot overlap.  For such systems temperature plays no role, so that the
partition function does not depend on $\beta$; we will write $y$ for the
fugacity of the new particles and $Q_\Lambda(y)$ for the corresponding
partition function.  It has been shown in particular for the case of dimers
(on an arbitrary graph) that the zeros of $Q_\Lambda(y)$ all lie on the
negative real $y$-axis \[HL]; any system with this property will of course
not have any phase transitions in the thermodynamic limit.  On the other
hand, hard diamonds on $\bbz^2$ and hard hexagons on the triangular lattice
do have a phase transition in the thermodynamic limit \[D,Bax,J].

In this note we will first describe a new class of Ising systems for which
no zeros touch the positive real axis, at least for large $\beta$ (low
temperature).  In some of these systems all the zeros lie on the negative
real axis, either for all values of $\beta$ or for large $\beta$; in
others, the zeros are excluded from some wedge $-\phi<\arg z<\phi$, where
$0<\phi<\pi$.  We will then investigate the ``hard'' systems obtained from
some of these, after suitable rescaling, in the limit $\beta\to\infty$;
these systems will similarly have no zeros of $Q(y)$ encroaching on the
real $y$-axis and hence no phase transition.  The models obtained in this
way include the monomer-dimer model \[HL] and the graph-counting models of
\[Ruelle3,Ruelle3a]; our results thus generalize these latter results to a
wider class of ``hard'' systems and to related low-temperature models.

\newsection{models} A class of systems with Lee-Yang zeros bounded away
from the positive real axis

We consider Ising spin systems decomposable into subsystems, called {\it
units}, with the property that each site belongs to either one or two
units.  Examples include the (three dimensional) pyrochlore lattice
\[MC1,MC2], in which the units are tetrahedra, the (two dimensional) kagome
lattice (\cfig{lattices}(a)), in which the units may be taken to be either
the triangles or the hexagons, the {\sl checkerboard} \[LS,LR], in which the
units are the alternate squares of the two dimensional square lattice
(\cfig{lattices}(b)), and the {\sl ladder}, in which every square is a unit
(\cfig{lattices}(c)).  We use the notation of \csec{intro} and write
$\Lambda_\alpha$ for the set of sites of the $\alpha^{\rm th}$ unit, with
$|\Lambda_\alpha|=n_\alpha$, and $\usigma_\alpha$ for the spin
configuration and $N_\alpha=N_\alpha(\usigma_\alpha)=N_\alpha(\usigma)$ for
the total number of up spins in the $\alpha^{\rm th}$ unit.  Note that in
general $N(\usigma)\le\sum_\alpha N_\alpha(\usigma)\le 2N(\usigma)$, since
a site $i$ with $\sigma_i=1$ may belong to either one or two units.

 \midinsert
\centerline{\epsfxsize=6.2truein
   \vbox{\hsize6.2truein\null\medskip\line{\hss\epsffile{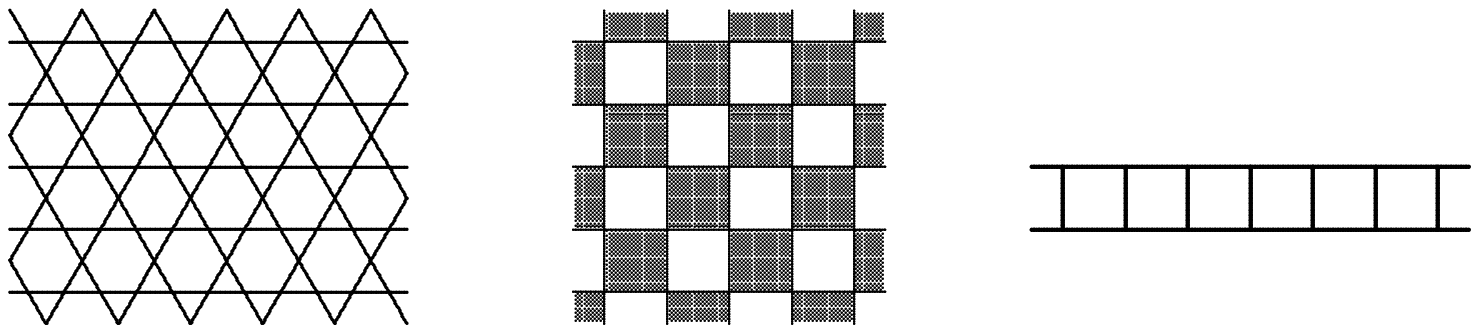}\hss}
 \bigskip
\line{\hss(a) Kagome lattice\hss\hss\hss\hss 
  (b) Checkerboard\hss\hss\hss\hss\hss
  (c) Ladder\hss\hss\hss}
\medskip
\centerline{\fig{lattices} Lattices decomposable into units with each site
 in two units.}}}
\endinsert

 The internal energy $U(\usigma)$ of the system is assumed to be the sum of
the internal energies of the units,
 $$U(\usigma)  = \sum_\alpha U_\alpha(\usigma_\alpha),\(edef)$$
  and these are assumed to be symmetric in the spins of the unit, so that
 $$U_\alpha(\usigma) =  F_\alpha(N_\alpha(\usigma)),\(eadef)$$
 with $F_\alpha$ a polynomial of degree at most $n_\alpha$.  We can think
of \(eadef) as a mean field interaction among the spins in
$\Lambda_\alpha$.  The partition functions \(zdef) for a single unit and
for the entire system thus become
 $$\eqalignno{
  Z_{\Lambda_\alpha}(z,\beta) 
   &= \sum_{\sigma_i=\pm1,\;i\in\Lambda_\alpha}
    z^{N_\alpha(\usigma_\alpha)}  
   e^{-\beta F_\alpha(N_\alpha(\usigma_\alpha))}
  =\sum_{\l=0}^{n_\alpha}{n_\alpha\choose \l}
    z^\l e^{-\beta F_\alpha(\l)},&\(zla)\cr
 Z_\Lambda(z,\beta) 
   &= \sum_{\sigma_i=\pm1}
    z^{N(\usigma)}  e^{-\beta \sum_\alpha F_\alpha(N_\alpha(\usigma))}.
  &\(zl)\cr}
  $$

We will prove in \csec{oneunit} that under certain conditions on the
function $F_\alpha$ the Lee-Yang zeros of $Z_\Lambda(z,\beta)$ are bounded
away from the positive real $z$-axis at low temperature or, under stronger
conditions, must lie on the negative real axis.  The next result shows that
such bounds on the zeros of $Z_\Lambda(z,\beta)$ follow from similar bounds
on the zeros of $Z_{\Lambda_\alpha}(z,\beta)$.

\thm{main} {\sl Suppose that the angle $\phi$ satisfies 
$0\le\phi<\pi/2$.  If  each zero $\zeta$ of 
$Z_{\Lambda_\alpha}(z,\beta)$ satisfies
 $$\zeta\ne0,\qquad \pi-\phi\le\arg \zeta \le\pi+\phi,\(sector3)$$
 then each zero $\zeta'$ of $Z_\Lambda(z,\beta)$  satisfies
 $$\zeta'\ne0,\qquad \pi-2\phi\le\arg \zeta' \le\pi+2\phi.\(sector4)$$
  In particular, if each $\zeta$ is real and negative, so is each
$\zeta'$.}
 
 \medskip
 Before proceeding to the proof of the theorem we give a simple example.

\ex{simple} If unit $\alpha$ has antiferromagnetic pair interactions of
equal strength between every pair of sites then its internal energy may be
written, after adding a constant, as
 $$F_\alpha(N_\alpha(\usigma_\alpha))
=-J\sum_{\{i,j\}\subset\Lambda_\alpha,\;i\ne j}(\sigma_i\sigma_j-1)
=-2\,|J|\,N_\alpha(n_\alpha-N_\alpha),\qquad J<0.\(simple)$$
 We consider several cases in which each unit has energy of the form \(simple).
 \smallskip\noindent
 (a) The one-dimensional nearest neighbor antiferromagnetic Ising model,
defined on $\Lambda=\{1,\ldots,n\}$, may be regarded as a model of this
type in which the units are the pairs $\alpha_i=\{i,i+1\}$, all with the
same coupling $J$.  From \(simple) we then have
$Z_{\Lambda_{\alpha_i}}(z,\beta)=z^2+2az+1$ with $a=e^{2\beta|J|}>1$;
this polynomial has two negative real zeros and hence, by \cthm{main}, the
zeros $Z(z,\beta)$ are negative real.

 \smallskip\noindent
 (b) When the only interactions considered for the checkerboard of
\cfig{lattices}(b) are pair interactions then one may term the system the
{\sl pyrochlore checkerboard} \[LS], since if each pair of vertices in a
square are connected with edges one obtains a planar representation of a
tetrahedron.  To be concrete we choose, for example, a $2L\times2L$ lattice
with doubly periodic boundary conditions.  It is then possible to check
that with the unit energy \(simple), with of course $n_\alpha=4$, all four
zeros of $Z_{\Lambda_\alpha}(z,\beta)$ are on the negative real axis (this
is in fact verified for arbitrary values of $n_\alpha$ in \cthm{SC1}
below).  \cthm{main} then states that the zeros of $Z_\Lambda$ will all lie
on the negative real axis; since this remains true as
$\Lambda\nearrow\bbz^2$, the system will not have a phase transition at any
finite temperature.  In fact the pressure and all correlations will be
analytic functions of $h$ for all $h\in\bbr$.

 \smallskip\noindent
 (c) The ladder (\cfig{lattices}(c)) illustrates the fact that two units
may share several vertices and thus an edge.  Note, however, that the
form \(eadef) of the total energy implies,with \(simple), that the coupling
constant for these shared edges (vertical in \cfig{lattices}(c)) is twice
that for the unshared (horizontal) edges. 

\subsection{proofmain} Proof of \cthm{main}

  The proof of \cthm{main} depends on two standard results, which we quote
for completeness; see the Appendix of \[R3] for more details.  We let
${\cal A}_n$ denote the space of complex polynomials in $z_1,\ldots,z_n$
which are separately affine in each variable, and observe that if $P$ is a
complex polynomial of degree at most $n$ then there is a unique symmetric
$\hat P\in{\cal A}_n$ such that $\hat P(z,\ldots,z)=P(z)$. A {\sl closed
circular region} is a closed subset $K$ of $\bbc$ bounded by a circle or a
straight line.

\lthm{Grace}{(Grace's theorem)} {\sl Let $P$ be a complex polynomial in one
variable of degree at most $n$.  If the $n$ roots of $P$ are contained in a
closed circular region $K$ and $z_1\notin K,\ldots,z_n\notin K$, then
$\hat P(z_1,\ldots,z_n)\ne0$.}

 \medskip
 If $P$ is in fact of degree $k$ with $k<n$ then we say that $n-k$ roots of
$P$ lie at $\infty$ and take $K$ noncompact.  For a proof of the result see
Polya and Szeg\"o \[PSz] V, Exercise 145.

\llem{ARlem}{(Asano-Ruelle) \[A,R2]} {\sl Let $K_1,K_2$ be closed
subsets of $\bbc$, with $K_1,K_2\not\ni0$.  If $\Phi$ is separately
affine in $z_1$ and $z_2$, and if
 $$	\Phi(z_1,z_2)\equiv A+Bz_1+Cz_2+Dz_1z_2\ne0      $$
whenever $z_1\notin K_1$ and $z_2\notin K_2$, then
  $$\tilde\Phi(z)\equiv A+Dz\ne0 $$
whenever $z\notin-K_1\cdot K_2$.  {\rm [}We have written
$-K_1\cdot K_2=\{-uv:u\in K_1,v\in K_2\}${\rm ]}.}

\medskip
 The map $\Phi\mapsto\tilde\Phi$ is called {\it Asano contraction}; we
denote it by $(z_1,z_2)\to z$. To state the next result we define, for
$\epsilon>0$ and $-\pi/2<\theta<\pi/2$,
$ K_\theta(\epsilon) =\{z:{\rm Re}[e^{i\theta}(z+\epsilon)]\le0\}$.

\lem{DR1a} {\sl If $\phi$ is as in \cthm{main} and $P$ is a complex
polynomial each of whose zeros $\zeta$ satisfies \(sector3),
then for any $\theta$ with $|\theta|<\pi/2-\phi$ there is an $\epsilon>0$
such that $\hat P(z_1,\ldots,z_n)\ne0$ when
$ z_1,\ldots,z_n\notin K_\theta(\epsilon)$.}

 \proof Clearly there is an $\epsilon>0$ such that $P(z)\ne0$ when
$ z\notin K_\theta(\epsilon)$, and the result follows from Grace's
theorem.\qed

\lem{DR1} {\sl Suppose that $\phi$ is as in \cthm{main} and that $P_i(z)$,
$i=1,\ldots,I$, is a polynomial of degree $n_i$ each of whose zeros $\zeta$
satisfies \(sector3).  Suppose further that the polynomial
$\hat Q(z_1,\ldots,z_n)$ is obtained from the product
 $$\prod_{i=1}^I\hat P_i(z_{i,1},\ldots,z_{i,n_i})$$
by a sequence of Asano contractions $(z_{i,j},z_{k,l})\to z_m$ or
relabelings $z_{i,j}\to z_m$.  Then each zero $\zeta'$ of $Q(z)$ satisfies
\(sector4).}

\proof For each $\theta$ with $|\theta|<\pi/2-\phi$ we obtain from
\clem{DR1a} and the Asano-Ruelle lemma that for some $\epsilon>0$,
$\hat Q(z_1,\ldots,z_n)\ne0$ when each of $z_1,\ldots,z_n$ lies in the
complement of the set $- K_\theta(\epsilon)\cdot K_\theta(\epsilon)$.  Thus
$Q(z)\ne0$ when $z$ is in the complement of
$- K_\theta(\epsilon)\cdot K_\theta(\epsilon)$.  This complement is the
interior of a parabola with focus at $0$ and in particular contains the ray
making an angle $2\theta$ with the positive real axis.  As $\theta$ varies
in $\theta\in(-\pi/2+\phi,\pi/2-\phi)$ this ray sweeps out the complement
of the region defined by \(sector4).\qed

 \medskip
 We can now give the proof of the main result.

\proofof{\cthm{main}} The main statement of the theorem is an immediate
consequence of \clem{DR1}, since $\hat Z(z_1,\ldots,z_{|\Lambda|})$ is
obtained from
$\prod_\alpha\hat Z_\alpha(z_{\alpha1},\ldots,z_{\alpha n_\alpha})$ by
Asano contractions and relabelings.  The last statement follows by taking
$\phi=0$. \qed

 \medskip
 In \cssec{quadratic} below we will need the following  corollary  of
 \clem{DR1}. 

\cor{DR1b} {\sl Suppose that $P(z)$ is a polynomial of degree $n$ for which
each zero $\zeta$ is real and negative. If $\hat Q$ is obtained from
$\hat P$ by squaring all coefficients, then each zero $\zeta'$ of $Q$ is real
and negative.}

\proof Take $\phi=0$, $I=2$ and $P_1=P_2=P$ in \clem{DR1} and make all
contractions $(z_{1j},z_{2j})\to z_j$.\qed

 \newsection{oneunit} Zeros of the partition function of a single unit

 In this section we consider a particular unit $\alpha$ with $n_\alpha$
sites, energy $F_\alpha(N_\alpha)$, and partition function
$Z_{\Lambda_\alpha}$, and address the question implicitly raised by
\cthm{main}: when are all zeros of the function $Z_{\Lambda_\alpha}$
confined to a sector of the form \(sector3) for some $\phi$?  In
\cssec{quadratic} we give a criterion which guarantees that for all $\beta$
these zeros satisfy \(sector3) with $\phi=0$, and in \cssec{convex} several
criteria implying bounds of the form \(sector3) for various values of $\phi$. 

\subsection{quadratic} A quadratic interaction energy

\thm{SC1} {\sl Suppose that $F_\alpha(\l)$ is quadratic with positive leading
coefficient: $F_\alpha(\l)=a\l^2+b\l+c$ with $a>0$.  Then for any $\beta\ge0$
all zeros of $Z_{\Lambda_\alpha}(z,\beta)$ are real and negative.}

 \medskip
 Note that if $F_\alpha(\l)=a\l^2+b\l+c$ then the constant $c$ is
irrelevant, the constant $b$ represents a shift in the magnetic field, and
the constant $a$ may be absorbed into the inverse temperature; thus we may
(and will) assume without loss of generality that $F_\alpha(\l)=-\l(n-\l)$.
In the spin language this is an energy in which every pair of spins in the
unit is coupled with the same antiferromagnetic interaction and there is a
uniform magnetic field, as in \cex{simple}; in the lattice gas language
particles on each pair of sites interact with the same positive repulsive
potential and there is a uniform chemical potential.

\lem{DR2}  {\sl If $b>0$, the polynomial
 $$P_{(b)}(z)=\sum_{\l=0}^n{n\choose \l}(1+b \l(n-\l))z^\l $$
has only real negative zeros.}

\proof We have 
 $$\eqalign{P_{(b)}(z)
  &=(z+1)^n+b n(n-1)\sum_{\l=1}^{n-1}{n-2\choose \l-1}z^\l\cr
  &=(z+1)^n+b n(n-1)z(z+1)^{n-2}\cr
 &=(z+1)^{n-2}[z^2+(2+b n(n-1))z+1],\cr}$$
which has only real negative zeros.\qed

\proofof{\cthm{SC1}} Starting from $P_{(2^{-k}\beta)}$ as in \clem{DR2}, we
obtain by $k$ applications of \ccor{DR1b} that the polynomial
 $$ \sum_{\l=0}^n{n\choose \l}(1+\beta \l(n-\l)2^{-k})^{2^k}z^\l  \(3.9a) $$
has only real negative zeros.  Letting $k\to\infty$ we find that
 $$ Z_{\Lambda_\alpha}(z,\beta)
    =\sum_{\l=0}^n{n\choose \l}e^{\beta \l(n-\l)}z^\l   \(3.9b)$$
has only nonpositive real zeros, and we need only observe that the constant
term in \(3.9b) is nonzero.  \qed

\rmk{aff} (a) If we consider the system to be comprised of a single unit,
i.e., take $\Lambda_\alpha=\Lambda$, then \cthm{SC1} implies that in the 
mean-field Ising model with antiferromagnetic interactions all Lee-Yang
zeros lie on the negative real axis.
 \smallskip\noindent
 (b) If we consider this same system but with ferromagnetic pair
interactions, which is equivalent to taking $\beta<0$ in \(3.9b), then the
standard Lee-Yang theory implies that all zeros of $Z_\Lambda$ lie on the
unit circle.

\subsection{convex} A convex interaction energy

When $F_\alpha$ is as in \cthm{SC1} it is convex on the range
$0\le \l\le n_\alpha$ in the sense that
 $$2F_\alpha(\l) \le F_\alpha(\l+1)+F_\alpha(\l-1),\quad 
    \l=1,\ldots,n_\alpha-1.\(cvx)$$
 In this section we consider a unit energy $F_\alpha(\l)$, not
necessarily quadratic, which satisfies \(cvx).

We begin by introducing some notation to describe such an $F_\alpha$ more
precisely.  Let $0=k_0<k_1<\cdots<k_{r-1}<k_r=n_\alpha$ be indices such
that strict inequality holds in \(cvx) if and only if $\l=k_i$ for some $i$
with $1\le i\le r-1$.  To understand the role of these indices it is
helpful to introduce a geometric interpretation.  Let $f_\alpha(x)$ be
defined on the interval $[0,n_\alpha]$ as the linear interpolation of the
nodes $(l,F_\alpha(l))$, $l=0,1,\ldots,n_\alpha$, and let
$f^*_\alpha\subset\bbr^2$ be the {\sl epigraph} of $f_\alpha$:
$f^*_\alpha=\{(x,y)\mid x\in[0,n_\alpha],\ y\ge f_\alpha(x)\}$.  Then
$f_\alpha^*$ is a convex subset of $\bbr^2$ with two vertical faces and $r$
nonvertical faces.  The vertices of $f^*_\alpha$ are the nodes
$(k_i,F_\alpha(k_i))$; all other nodes $(l,F_\alpha(l))$ are interior
points of the (nonvertical) faces of $f^*_\alpha$.  See \cfig{poly}.  For
$1\le i\le r$ we define $H_{\alpha,i}$ to be the slope of the (nonvertical)
face of $f_\alpha^*$ containing $(k_{i-1},F_\alpha(k_{i-1}))$ and
$(k_i,F_\alpha(k_i))$, and note that $H_{\alpha,i}=F_\alpha(\l)-F_\alpha(\l-1)$
whenever $k_i-1<\l\le k_i$.  Finally, for $h\in\bbr$ we define
 $$E_\alpha(h)=\min_{0\le \l\le n}(F_\alpha(\l)-h\l).\(minE)$$
 We will be interested in \csec{ground} in the set $S_\alpha(h)$ of values
of $l$ on which the minimum in \(minE) is realized; clearly if $h$ is not
equal to any of the $H_{\alpha,i}$ then $S_\alpha(h)$ contains a unique
$l$, while for $h=H_{\alpha,i}$ it contains those $\l$ for which
$(\l,F_\alpha(l))$ lies in the $i^{\rm th}$ nonvertical face of
$f^*_\alpha$.

The next result shows that at low temperature the zeros of
$Z_{\Lambda_\alpha}$ fall into $r$ groups, where the $i^{\rm th}$ group is
naturally associated with the $i^{\rm th}$ nonvertical face of $f^*_\alpha$
and contains $k_i-k_{i-1}$ points, all with magnitude of order
$e^{-\beta H_{\alpha,i}}$.

\midinsert
\centerline{\epsfxsize=3.87truein
   \vbox{\hsize3.87truein\line{\hss\epsffile{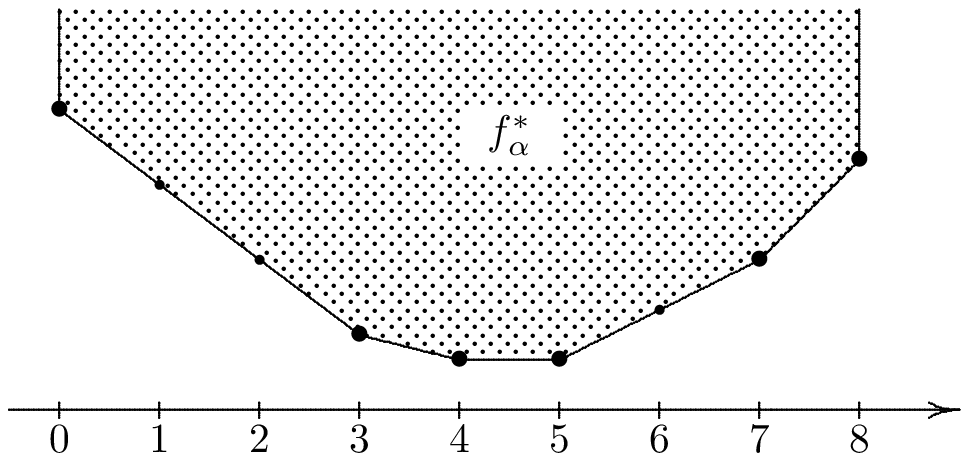}\hss}
 \bigskip
 \centerline{\fig{poly} Typical  set $f^*_\alpha$, with
 $n_\alpha=8$, $r=5$, and $(k_0,\ldots,k_5)=(0,3,4,5,7,8)$.}
 \smallskip
 \centerline{Nodes $(l,F_\alpha(l))$ are indicated by dots, with heavier
 dots when $l=k_i$ for some $i$.}}}
\endinsert

\lem{SC2} {\sl  For $i=1,\ldots r$ let
 $$R_i(t)=\sum_{j=k_{i-1}}^{k_i}{n_\alpha\choose j}t^{j-k_{i-1}}$$
 and let $t_{i,1}\ldots t_{i,k_i-k_{i-1}}$ be the zeros of $R_i$.  Then one
may number the zeros $z_1,\ldots z_{n_\alpha}$ of $Z_{\Lambda_\alpha}$ in
such a way that for $k_{i-1}<j\le k_i$,
 $$\lim_{\beta\to\infty} z_je^{\beta H_{\alpha,i}}=t_{i,j-k_{i-1}}.$$}

 \proof For $\l< k_{i-1}$ or $\l>k_i$ the coefficient of $t^\l$ in the
polynomial
 $$\eqalign{\hat R_{\beta,i}(t)
  &=e^{\beta E_\alpha(H_{\alpha,i})}
             Z_{\Lambda_\alpha}(te^{\beta H_{\alpha,i}},\beta)\cr
  &=t^{k_{i-1}}R_i(t)+\sum_{\l<k_{i-1}\;{\rm or\;\l>k_i}}
   {n_\alpha\choose \l}t^\l e^{-\beta (F_\alpha(\l)-H_{\alpha,i}\l
        -E_\alpha(H_{\alpha,i}))}
  \cr}$$
 converges to zero as $\beta\nearrow\infty$.  Thus $k_{i-1}$ of the roots
converge to 0, $n_\alpha-k_i$ to infinity, and the remaining $k_i-k_{i-1}$
to the roots of $R_i$ \[C].\qed

 \medskip
  To state the main result of this section we let
$\delta=\max_{1\le i\le r}(k_i-k_{i-1})$; $\delta+1$ is the maximum number
of nodes lying on any nonvertical face of $f^*_\alpha$.

\thm{sec} {\sl (a) If $\delta=1$, i.e, if all the inequalities in \(cvx)
are strict, then all roots of $Z_{\Lambda_\alpha}$ are real and negative
for sufficiently large $\beta$.
 \smallskip\noindent
 (b) If $\delta=2$ then all roots of $Z_{\Lambda_\alpha}$ satisfy
 \(sector3) with $\phi=\pi/3$ for sufficiently large $\beta$.
 \smallskip\noindent
 (c) If $\delta=3$ then there is an angle $\phi_{n_\alpha}$, which as
indicated may be chosen to depend only on $n_\alpha$, such that
$\phi_{n_\alpha}<\pi/2$ and such that all roots of $Z_{\Lambda_\alpha}$
satisfy \(sector3) with $\phi=\phi_{n_\alpha}$ for sufficiently large
$\beta$.
 \smallskip\noindent
 (d) If $k_1\le4$ and $k_{r-1}\le n_\alpha-4$ with at least one of these an
equality, and $k_i-k_{i-1}\le3$ for $i-2,\ldots,r-1$ so that $\delta=4$,
then there is an angle $\phi_{n_\alpha}<\pi/2$ such that all roots of
$Z_{\Lambda_\alpha}$ satisfy \(sector3) with $\phi=\phi_{n_\alpha}$ for
sufficiently large $\beta$.}

\proof (a) When $\delta=1$ each of the polynomials $R_i(t)$ is linear, with
a negative real root.  From \clem{SC2} the roots of $Z_{\Lambda_\alpha}$
for large $\beta$ must  be widely separated in magnitude, and since
any complex roots among these occur in complex conjugate pairs, the roots
must in fact be real.

 \smallskip\noindent
 (b) It follows from \clem{SC2} that $\arg z_j\to\arg t_{i,j-k_{i-1}}$ as
$\beta\nearrow\infty$, so that it suffices to show that each root $t_{ij}$
satisfies $|\arg t_{ij}-\pi|<\pi/3$.  When $\delta=2$ the $R_i(t)$ are either
linear, with roots having argument $\pi$, or quadratic; in the latter case
the quadratic formula shows that the roots $t_{ij}$ are complex, have
negative real part, and satisfy
 $$\left|\Im t_{ij}\over \Re t_{ij}\right|
    =\sqrt{{4(k_i-1)(n_\alpha+1-k_i)\over k_i(n_\alpha+2-k_i)}-1}<\sqrt3,
    \(tan)$$
 which yields the desired bound. Improved bounds on the roots for specific
values of $k_i$ and $n_\alpha$ may be obtained from \(tan).  For example,
if $i=1$ then $k_1=2$ and one obtains \(sector3) with $\phi=\pi/4$ for the
smallest (in magnitude, at large $\beta$) two zeros of
$Z_{\Lambda_\alpha}$, a result closely related to earlier work of Ruelle
\[Ruelle3], as we discuss in \csec{graphpolys}.

 \smallskip\noindent
 (c) When $\delta=3$ the polynomials $R_i$ can be linear, quadratic, or cubic;
following the analysis of (b) it suffices to show that in the cubic case
all roots have negative real part.  Up to a constant factor any such cubic
$R_i$ has the form
 $$\eqalign{k(k-1)&(k-2)+k(k-1)(n+3-k)t\cr
   &+k(n+3-k)(n+2-k)t^2
   +(n+3-k)(n+2-k)(n+1-k)t^3,\cr}\(cubic)$$
 where $n=n_\alpha$ and $k=k_i$.  For $n=k=3$ this polynomial has a triple
root at $z=-1$; we may then vary $n$ and $k$ continuously to some desired
values and ask whether roots can cross the imaginary axis during this
procedure.  We may assume that the intermediate values of $n,k$ remain real
and satisfy $n,k\ge3$ and $n-k\ge0$.  Suppose (3.6) vanishes for $t=is$,
$s$ real.  We cannot have $s=0$ since $k(k+1)(k+2)\ne0$.  For $s\ne0$ we
would have both
 $$(k-1)(k-2)=(n-k+2)(n-k+3)s^2$$
 and
 $$k(k-1)=(n-k+1)(n-k+2)s^2,$$
 so that $(n-k+1)(k-2)=k(n-k+3)$, i.e., $n+1=0$, in contradiction with
$n\ge3$.

 \smallskip\noindent
 (d) We suppose that $k_1=4$; the analysis when $k_{r-1}=n_\alpha-4$ is the
 same.  With the results (a)--(c) above it suffices to show that the roots of

 $$R_1(t)=\sum_{\l=0}^4{n_\alpha\choose \l}t^\l\(R1)$$
 satisfy an appropriate bound of the form \(sector3). The roots of $R_1(t)$
are all equal to $-1$ for $n_\alpha=4$ and, treating $n_\alpha$ as a
continuous variable, can have the form $t=is$, $s$ real, only for
$n_\alpha$ a root of $n^2+9n-4=0$; as both the roots of this polynomial are
less than 4 the roots of $R_1$ for $n_\alpha>4$ must all lie strictly in
the left half plane.\qed

\rmk{newt} (a) A classical result of Newton provides a converse to
\cthm{sec}(a): if all roots of $Z_{\Lambda_\alpha}$ are real for some
$\beta$ then either equality holds for all $\l$ in \(cvx), in which case
$F_\alpha(\l)=A+B\l$ for some $A,B$, there are no interactions, and all roots
of $Z_{\Lambda_\alpha}$ are equal, or strict inequality holds for all $\l$
in \(cvx).  For a proof see \[HLP], page 104.  Other related results are
contained in \[K,H,BDN].  In particular, (a) of the theorem follows from the
result of \[K].
 \smallskip\noindent
 (b) By taking $k\approx n_\alpha/2$ one sees that there is no bound
\(sector3) on the roots of \(cubic) which is uniform in $n_\alpha$ and $k$
and satisfies $\phi<\pi/2$.  On the other hand, one can show that such a
uniform bound may be found  both for the roots of \(cubic)
for fixed $k$ and for the roots of \(R1).

 \subsection{fourspin} An example: quartic and quadratic  interactions

As an example we consider a unit with two and four spin interactions which
satisfy spin flip symmetry.  In the particle language described
in \csec{intro} the energy is
 $$U_\alpha(\usigma)
   =  -K_2\sum_{1\le i< j\le n}(\eta_i\eta_j+\hat\eta_i\hat\eta_j)
 -K_4\sum_{\textstyle{X\subset\{1,\ldots,n\}\atop|X|=4}}
    \left(\textstyle\prod_{i\in X}\eta_i
    +\textstyle\prod_{i\in X}\hat\eta_i\right),$$
 where $\hat\eta_i=1-\eta_i$, i.e.,
 $$ F_\alpha(\l)=-K_2\left[{\l\choose2}
     +{n_\alpha-\l\choose2}\right]
   -K_4\left[{\l\choose4}+{n_\alpha-\l\choose4}\right].$$
 We assume that $K_2$ and $K_4$ are not both zero.  The convexity condition
\(cvx) is satisfied with strict inequality for all $\l$ if
 $$2K_2+{K_4\over2}\,\bigl[(l-1)(l-2)+(n_\alpha-l-1)(n_\alpha-l-2)\bigr]<0$$
 for $l=1,\ldots,n-1$.  This happens when
 $$\theta_{n_\alpha}<\arg(K_2+iK_4)<\phi_{n_\alpha},\(niceK)$$
 where the angles $\phi_n$ and $\theta_n$ are given by
 $$\matrix{\tan\theta_n=-\ds{4\over(n-2)(n-3)},
    \hfill&\qquad \pi/2\le\theta_n<\pi,\hfill\cr
\noalign{\smallskip}
  \tan\phi_n=\cases{-\ds{8\over(n-2)(n-4)},&if $n$ is even,\cr
   -\ds{8\over(n-3)^2},&if $n$ is odd,\cr}   
   \hfill&\qquad 3\pi/2\le\phi_n<2\pi.\hfill\cr}$$
 See \cfig{plane1}.  Under condition \(niceK), \cthm{sec}(a) implies that
$Z_{\Lambda_\alpha}$ has its zeros on the negative real axis at low
temperature, and hence by \cthm{main} so does $Z_\Lambda$, if all units in
the system are of this type.  Note that in particular \(niceK) includes the
negative $K_2$-axis, where we know from \cthm{SC1} that the zeros are real
and negative at all temperatures.  For nonzero values of $K_2$ and $K_4$
not satisfying \(niceK), \crmk{newt}(a) implies that the zeros do not lie
exclusively on the real $z$-axis for any $\beta$.

 \midinsert
\centerline{\epsfxsize=2.125truein
   \vbox{\hsize2.125truein\line{\hss\epsffile{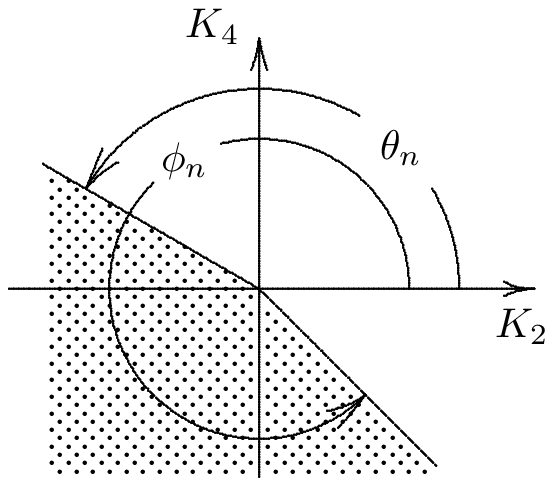}\hss}
 \medskip
 \centerline{\fig{plane1} In the shaded region (which extends to infinity
in both}
 \centerline{the $x$ and $y$ directions) all zeros are real and negative at
low temperature.}}}
\endinsert

 If we specialize further to the case $n_\alpha=4$, in which
$\theta_n=\tan^{-1}(-2)$ and $\phi_n=3\pi/2$, then the region in the space
of interactions at which all zeros are on the negative real axis can be
computed exactly.  Since a change of temperature is equivalent to a
rescaling of $(K_2,K_4)$ it is convenient to take $\beta=1$; then this
region is given by
 $$\log\left[3-\sqrt{9-8a^2}\over2a^4\right] > K_4
   > \cases{\ds\log\left[4a-3\over a^4\right],&if $K_2>\log(3/4)$,\cr
     -\infty,&otherwise,}$$
 where $a=e^{K_2}$.  See \cfig{plane2}.  The computation follows that in
the proof of Proposition 6 of \[LR], and we omit details.

 \midinsert
\centerline{\epsfxsize=4.0truein \epsfysize=3.2truein
   \vbox{\hsize4truein{\line{\hss\epsffile{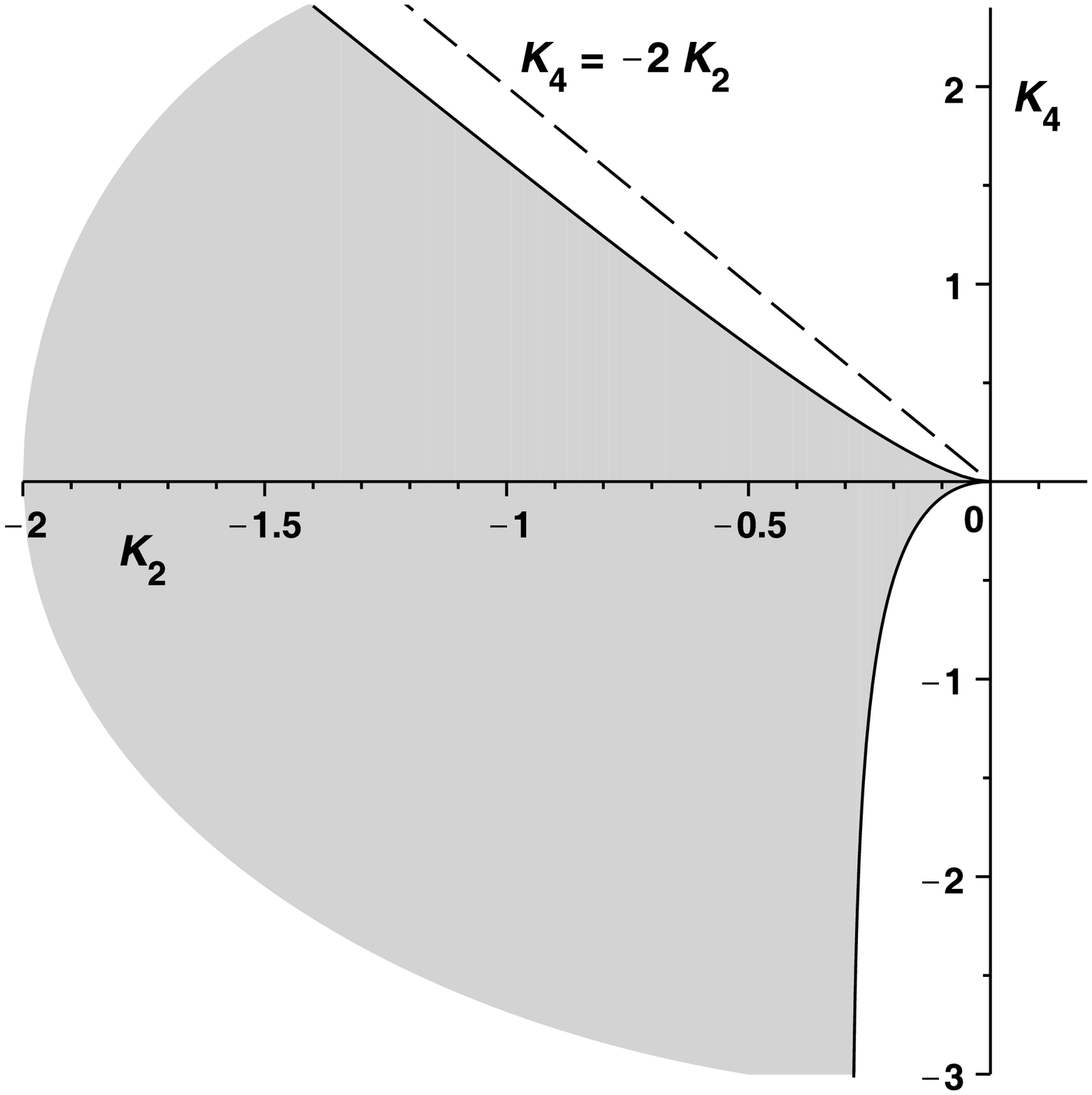}\hss}
 \medskip
 \centerline{\fig{plane2} The case $n_\alpha=4$. When $\beta=1$ all zeros
are real and negative} \centerline{if and only if $(K_2,K_4)$ lies in the
shaded region (which extends} \centerline{to infinity in both the $x$ and
$y$ directions).}}}}
\endinsert

\newsection{ground} Ground states and zero temperature limits

Consider a system which is assembled from units, as described in
\csec{models}, such that the energy $F_\alpha$ for each unit is convex in
the sense of \(cvx).  In this section we suppose further that each site
belongs to exactly two units.

Let us fix, for the moment, a magnetic field $h_0$.  The total energy of the
system in spin configuration $\usigma$, including the magnetic energy, is
then
 $$U(\usigma)-2h_0N(\usigma)
   =\sum_\alpha [F_\alpha(N_\alpha(\usigma)) -h_0N_\alpha(\usigma)].\(Esum)$$
 From \(Esum) and \(minE) it follows that this energy is bounded below by
$E_0=\sum_\alpha E_\alpha(h_0)$.  On the other hand, if we recall the
definition of $S_\alpha(h_0)$ given below \(minE) we see that $E_0$ is in
fact the ground state energy of the system---the minimum value of
\(Esum)---if and only if it is possible to find a spin configuration
$\usigma$ such that $N_\alpha(\usigma)\in S_\alpha(h_0)$ for each $\alpha$.
When this is true we say that the system is {\sl not frustrated}.

Now we assume that our system is not frustrated and consider  a zero
temperature limit $\beta\nearrow\infty$ with a $\beta$-dependent fugacity
$z(\beta)=e^{2\beta h(\beta)}$ such that $h(\beta)\to h_0$ as
$\beta\nearrow\infty$; specifically, for some $\lambda\in\bbr$ we take
 $$h(\beta)=h_0+{\lambda\over2\beta}.\(slope)$$
 In the $h$-$T$ phase plane (where $T=1/\beta$ is the temperature) this
corresponds to approaching $(h_0,0)$ along a line with slope $2/\lambda$
\[GLM].  Then with $y=e^\lambda$ the limiting partition function is
 $$\eqalignno{Q_\Lambda(y,h_0)
  &=\lim_{\beta\nearrow\infty} e^{\beta E_0}Z_\Lambda(z(\beta),\beta)\cr
 &=\lim_{\beta\nearrow\infty}
  \sum_{\usigma}
    y^{N(\usigma)}  e^{-\beta (U(\usigma)-2h_0N(\usigma)-E_0)}
 =\sum_{\usigma\in\G(h_0)} y^{N(\usigma)},&\(Qdef)\cr}$$
  where $\G(h_0)$ is the set of ground-state configurations.  We are of
course interested in the behavior of the zeros of $Z_\Lambda$ under the
limiting process \(Qdef).  If
$N_{\rm min}(h_0)=\min_{\usigma\in\G(h_0)}N(\sigma)$ and
$N_{\rm max}(h_0)=\max_{\usigma\in\G(h_0)}N(\sigma)$ then $N_{\rm min}$
zeros will converge to 0 and $|\Lambda|-N_{\rm max}$ to $\infty$ \[C],
while the remaining zeros converge to the (nonzero) roots of
$y^{-N_{\rm min}}Q_\Lambda(y,h_0)$.

When for each $\alpha$ one has $|S_\alpha(h_0)|=1$, that is, when
$h_0\ne H_{\alpha,i}$ for any $\alpha,i$, there is a unique ground state
configuration and $Q(y,h_0)$ is rather uninteresting.  When there are many
ground state configurations, however, they can in some cases be identified
with configurations of ``hard objects'' and $Q(y,H_0)$ is then the
partition function for these.  

In the next example we illustrate these ideas by revisiting
\cex{simple}(b).  In \csec{graphpolys} we describe a family of examples
involving {\sl graph-counting polynomials}.

\ex{simplebis} We consider again \cex{simple}(b): pair interactions on a
$2L\times2L$ pyrochlore checkerboard with doubly periodic boundary
conditions.  The unit energy of the model is given in \(simple), with
$n_\alpha=4$ for all $\alpha$.  Since the energy for all units has the same
form $F_\alpha(l)=-2|J|l(4-l)$ we will omit the subscript $\alpha$ on $F$
and similar quantities when no confusion can arise.  The fields
$H_l=F(l)-F(l-1)$ defined in \cssec{convex} are $H_1=-6|J|$, $H_2=-2|J|$,
$H_3=2|J|$, and $H_4=6|J|$.  For $H_l<h_0<H_{l+1}$ (with $H_0=-\infty$ and
$H_5=\infty$) the zero temperature limit along the line \(slope) is
independent of $\lambda$ and the ground state configurations each have
exactly $l$ up spins in each unit, that is, $N_\alpha=l$ for each $\alpha$.
If we take the $T\to0$ limit of the partition function along a line
\(slope) with $h_0=H_l$ we obtain ground states in which both $N_\alpha=l$
and $N_\alpha=l-1$ are possible, with the total value of $N$ controlled by
the fugacity $y=e^\lambda$.  The situation in the $h$-$T$ plane is shown in
\cfig{pcht}, with a typical line \(slope) for $h_0=H_1=-6|J|$.

 \midinsert
\centerline{\epsfxsize=4.35truein
   \vbox{\hsize4.35truein\line{\hss\epsffile{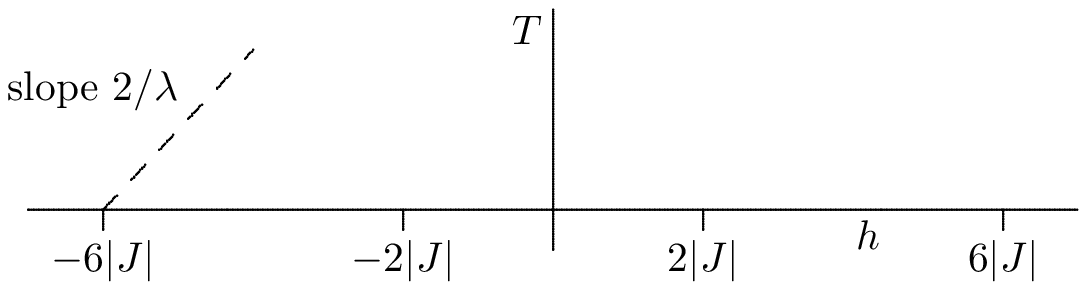}\hss}
\medskip
\centerline{\fig{pcht} The $h$-$T$ plane for the pyrochlore checkerboard
with pair interactions.}
}}
\endinsert

We may interpret the ground state configurations in terms of hard objects
by considering dimers on the new lattice $\Lambda'$ obtained from $\Lambda$
by shrinking each unit to a vertex and introducing an edge joining two of
these points when the corresponding units share a site; $\Lambda'$ is again
a square lattice with certain periodic boundary conditions.  An occupied
site in $\Lambda$ corresponds to a dimer covering the corresponding edge in
$\Lambda'$, so that a ground state for $H_l<h_0<H_{l+1}$, with with
$N_\alpha=l$ for each $\alpha$, corresponds to a dimer configuration on
$\Lambda'$ in which every vertex is covered by exactly $l$ dimers.  In a
ground state with $h_0=H_l$ each site of $\Lambda'$ is covered by either
$l$ or $l-1$ dimers. Thus, for example, for $h_0=H_1=-6|J|$ these are
monomer-dimer configurations; for $h_0=H_2=-2|J|$ they are a restricted
class of {\sl unbranched subgraphs} \[Ruelle3a].  In each case,
\cthm{sec}(a) implies that all zeros of the partition function lie on the
negative real axis, a result originally obtained for the monomer-dimer
system (in much more generality) in \[HL].

\newsection{graphpolys} Graph-counting polynomials

Consider a graph $G$ with sets $V$ of vertices and $E$ of edges, such that
each edge connects a pair of distinct vertices.  Note that such graphs may
have several edges joining the same pair of vertices.  We let $d_0$ be the
maximum vertex degree in $G$. A {\sl subgraph} $M$ of $G$ is a graph with
vertex set $V$ and edge set contained in $E$; $|M|$ denotes the number of
edges of $M$ and for any vertex $v\in V$ we let $d_{M}(v)$ denote the
vertex degree of $v$ in $M$ that is, number of edges of $M$ incident
on~$v$.

A {\sl graph-counting polynomial} \[Ruelle3] is a polynomial
 $$\tilde Q_{C}(y)=\sum_{M\in (C)}y^{|M|},$$
 where $(C)$ is the collection of subgraphs of $G$ associated with some set
$C$ of nonnegative integers via
$(C)=\{M\mid d_M(v)\in C,\ \forall v\in V\}$.  For $C=\{0,1\}$, $(C)$ is
the class of {\sl dimer subgraphs} of $G$; in this case it was shown by
Heilmann and Lieb \[HL] that all the zeros of $\tilde Q_{C}$ are real and
negative.  For $C=\{0,1,2\}$, $(C)$ is the set of {\sl unbranched
subgraphs} of $G$ and it is shown in \[Ruelle3a] that the zeros of
$\tilde Q_{C}$ lie in the left half plane; results for other choices of
$C$ are given in \[Ruelle3].  In this section we study $\tilde Q_{C}$ for
$C$ a nonempty interval $C_{pq}=\{p,p+1,\ldots,q\}$ of nonnegative
integers.

Given the graph $G$ and the set $C_{pq}$ we introduce a statistical
mechanical system of the type described in \csec{models}.  In this system
$\Lambda=E$; we may think of starting with a geometric realization of the
graph and then putting a site of $\Lambda$ at the center of each edge.  For
each $v\in V$ there is a unit $\alpha_v$ which contains those sites of
$\Lambda$ which correspond to edges of $G$ incident on $v$.  Since in $G$
each edge is incident on two vertices, this system has the property,
assumed in \csec{ground}, that every site belongs to exactly two units.
Finally, we introduce a unit energy $F(l)$, $l=0,1,\ldots,d_0(G)$, the same
for all units, which satisfies \(cvx) and is such that, with $k_i$ the
indices defined in \cssec{convex}, (i)~$p=k_{i-1}$ and $q=k_i$ for some
$i$, and (ii)~$k_j=k_{j-1}+1$ for all $j\ne i$.  In other words, the
$i^{\rm th}$ nonvertical face of the convex set $f^*$ associated with $F$
(see \cfig{poly}) contains the nodes $(l,F(l))$ for $l=p,\ldots,q$, and all
other faces contain exactly two nodes.

Now consider the limit \(Qdef) with $h_0=H_i=(F(q)-F(p))/(q-p)$.  The
ground state configurations at magnetization $H_i$ are precisely those in
which $N_{\alpha_v}\in C_{pq}$ for each $v$, and the assumption that the
system is not frustrated is precisely the assumption that such
configurations exist.  Each such configuration, however, has an immediate
interpretation as a subgraph of $G$: an edge $e\in E$ belongs to the
subgraph if and only if the corresponding site in $\Lambda$ is occupied
(using the lattice gas language).  With this identification the ground
state configurations then give rise precisely to the subgraphs belonging to
$(C_{pq})$, and the limiting partition function $Q_\Lambda(y,H_i)$ is the
same as the graph-counting polynomial $\tilde Q_{C_{pq}}(y)$.

The next result, which describes the behavior of the zeros of
$Q_{C_{pq}}$ for certain choices of $p,q$, follows immediately from 
\cthm{sec}.

\thm{sec2} {\sl Suppose that $p$ and $q$ are such that $(C_{pq})$ is
nonempty.  Then:
 \smallskip\noindent
 (a) If $q=p+1$ then all the nonzero roots of $\tilde Q_{C_{pq}}(y)$ are
real and negative.
 \smallskip\noindent
 (b) If $q=p+2$ then all the nonzero roots of $\tilde Q_{C_{pq}}(y)$
satisfy \(sector4) with $\phi=\pi/3$.
 \smallskip\noindent
 (c) If $q=p+3$ then there is an angle $\phi$, which may depend on $p$,
$q$, and $d_0(G)$ but not on the size of the graph, such that $\phi<\pi/2$
and such that all the nonzero roots of $\tilde Q_{C_{pq}}(y)$ satisfy
\(sector4) with the angle $\phi$.
 \smallskip\noindent
 (d) If $p=0$ and $q=4$ or $p=d_0(g)-4$ and $q=d_0(G)$ then there is an
angle $\phi<\pi/2$, which depends on $d_0(G)$ but may be chosen uniformly
in the size of $G$, such that all the nonzero roots of
$\tilde Q_{C_{pq}}(y)$ satisfy \(sector4) with the angle $\phi$.}

 \bigskip\noindent
 {\bf Acknowledgments.}  We thank Joseph Slawny for helpful conversations.
 The work of JLL  was supported in part by NSF Grant  DMR-1104501 and AFOSR
grant FA9550-10-1-0131.

\references

\medskip\noindent
 \[Ruellebook] D. Ruelle, {\sl Statistical Mechanics: Rigorous Results}.
World Scientific, Singapore,  1999.

 \smallskip\noindent
\[YL] C. N. Yang and T. D. Lee,  Statistical theory of equations of state
and phase transitions I:  Theory of condensation.  {\sl Phys. Rev.} {\bf
87}, 404--409 (1952).

 \smallskip\noindent
 \[LY] T. D. Lee and C. N. Yang,  Statistical theory of equations of state
and phase transitions II:  Lattice gas and Ising model.  {\sl Phys. Rev.} {\bf
87}, 410--419 (1952).

 \smallskip\noindent
 \[SF] M. Suzuki and M. E. Fisher, Zeros of the partition function for the
 Heisenberg, ferroeletric, and general Ising models.  {\sl J. Math. Phys.}
{\bf 12}, 235--246 (1971).

 \smallskip\noindent
 \[BBCKK] M. Biskup, C. Borgs, J. T. Chayes, L. J. Kleinwaks, and R.
Koteck\'y, General theory of Lee-Yang zeros in models with first-order
phase transitions.  {\sl Phys. Rev. Lett.} {\bf 84}, 4794--4797 (2000). 

 \smallskip\noindent
  \[Kim] S.-Y. Kim, Yang-Lee zeros of the antiferromagnetic Ising model.  
{\sl Phys. Rev. Lett.} {\bf 93}, 130604 1--4 (2004).

 \smallskip\noindent
  \[HK] C.-O. Hwang and S.-Y. Kim, Yang-Lee zeros of triangular Ising
antiferromagnets.  {\sl Physica A} {\bf 389}, 5650--5654 (2010).
 
 \smallskip\noindent
 \[LR1] E.H. Lieb and D. Ruelle, A property of the zeros of the partition
function for Ising spin systems.  {\sl J. Math. Phys.} {\bf
13}, 781-784 (1972).

 \smallskip\noindent
  \[D] R. L. Dobrushin, The problem of uniqueness of a Gibbsian random
field and the problem of phase transition. {\sl Funct. Anal. Appl.} {\bf 2},
302--312 (1968).

 \smallskip\noindent
  \[DKS] R. L. Dobrushin, J. Kolafa, and S. B. Shlosman, Phase diagram of
the two-dimensional Ising antiferromagnet (computer-assisted proof).
{\sl Commun. Math. Phys.} {\bf102}, 89--103 (1985).

 \smallskip\noindent
  \[Sin] Ya. G. Sinai., {\sl Theory of phase transitions: rigorous results}. 	 New York, Pergamon Press, 1982.

 \smallskip\noindent
 \[LR] Joel L. Lebowitz and David Ruelle, Phase transitions with four-spin
interactions. {\sl Commun. Math. Phys.} {\bf 304}, 711--722 (2011).

 \smallskip\noindent
  \[HL] O. J. Heilmann and E. H. Lieb, Theory of monomer-dimer systems.
  {\sl Commun. Math. Phys.} {\bf 25}, 190--232 (1972).

 \smallskip\noindent
 \[Bax] R. J. Baxter, Hard hexagons: exact solution. {\sl J. Phys. A:
 Math. Gen.} {\bf 13} L61-L70 (1980).

 \smallskip\noindent
  \[J] G. S. Joyce, On the hard-hexagon model and the theory of modular
 functions.  {\sl Phil. Trans. R. Soc. Lon.} {\bf 325}, 643--702 (1988).

 \smallskip\noindent
 \[Ruelle3] David Ruelle, Zeros of graph-counting polynomials. {\sl Journ.
Algebr. Comb.} {\bf 9}, 157--160 (1999).

 \smallskip\noindent
 \[Ruelle3a] David Ruelle, Counting unbranched subgraphs. {\sl Commun.
Math. Phys.} {\bf 200}, 43--56 (1999).

 \smallskip\noindent
 \[MC1] R. Moessner and J. T. Chalker, Properties of a classical spin
liquid: the Heisenberg pyrochlore ferromagnet. {\sl Phys. Rev. Lett.} {\bf
80}, 2929--2932 (1998).

 \smallskip\noindent
 \[MC2] R. Moessner and J. T. Chalker, Low-temperature properties of
classical geometrically frustrated antiferromagnets.  {\sl Phys. Rev. B}
{\bf 58}, 12049--12062 (1998).

 \smallskip\noindent
 \[LS] Elliott H. Lieb and Peter Schupp, Ground state properties of a fully
 frustrated quantum spin system. {\sl Phys. Rev. Lett.} {\bf 83},
 5362--5365 (1999).

 \smallskip\noindent
 \[R3] D. Ruelle, Characterization of Lee-Yang polynomials.  {\sl Ann. Math.}
{\bf 171}, 589--603 (2010).

 \smallskip\noindent
 \[PSz]  G. P\'olya and G. Szeg\"o,  {\it Problems and theorems in
analysis II.} Springer, Berlin, 1976.

  \smallskip\noindent
 \[A] T. Asano, Theorems on the partition functions of the Heisenberg
ferromagnets. {\sl J. Phys. Soc. Jap.} {\bf 29}, 350--359 (1970).

 \smallskip\noindent
 \[R2] D. Ruelle, Extension of the Lee-Yang circle theorem.  {\sl Phys. Rev.
Lett.} {\bf 26}, 303--304 (1971).

 \smallskip\noindent
  \[C] J. L.  Coolidge, The continuity of the roots of an algebraic equation.
{\sl Ann. Math.} {\bf 9}, 116-118 (1908).

 \smallskip\noindent
  \[HLP] G. H. Hardy, J. E. Littlewood, and G. P\'olya, {\it Inequalities}.
  Cambridge University Press, Cambridge, 1952.

  \smallskip\noindent
 \[K] Kurtz, D.C., A sufficient condition for all the roots of a
polynomial to be real. {\sl Amer. Math. Monthly} {\bf 99},259--263 (1992).

 \smallskip\noindent
 \[H] Handelman, D., Arguments of zeros of highly log concave polynomials.
arXiv:1009. 6022v1 [math.CA].

 \smallskip\noindent
 \[BDN] P. Bahls, R. Devitt-Ryder, and T. Nguyen, The location of roots of
logarithmically concave polynomials. Preprint; submitted to Ann. Comb.

 \smallskip\noindent
 \[GLM] G.Giacomin, J. L. Lebowitz, and C.Maes, Agreement percolation and
phase coexistence in some Gibbs systems. {\sl Journal of Statistical Physics},
{\bf 80}, 1379-1403 (1995).

\bye